\documentclass[prd,nobibnotes,nofootinbib,showpacs,10pt]{revtex4}
\pdfoutput=1
\usepackage[latin1]{inputenc}
\usepackage{graphicx}
\usepackage{amsmath}

\usepackage{amssymb}
\usepackage{braket}
\DeclareMathAlphabet{\mathpzc}{OT1}{pzc}{m}{it}
\DeclareMathOperator{\Tr}{\mathbf{Tr}}
\DeclareMathOperator{\Op}{\mathcal{O}}
\numberwithin{equation}{section}
\usepackage[mmddyyyy]{datetime}
\newdate{date}{23}{06}{2021}

\begin{document}
	\title{Deconstructing finite temperature pure gauge theory.}
	\author{Herbert Neuberger}
	\email{herbert.neuberger@gmail.com}
	\affiliation{Department of Physics and Astronomy, Rutgers University,\\ 
		Piscataway, NJ 08854, U.S.A} 
	
	\begin{abstract}
	A deconstructed finite temperature gauge theory has the Euclidean time direction kept discrete and finite. The ultraviolet behavior is that of one dimension less than at zero temperature. One can add to the action a gauge invariant term dependent on Polyakov loops, without worrying about nonlocality. The deformation of {\" U}nsal and Yaffe is best analyzed in this
	framework. It is shown that turning it on causes a dramatic change in the Hilbert space on which the transfer matrix in the temperature direction acts.
	The undeformed Hilbert space is everywhere a gauge singlet while the deformed one includes all states transforming locally, anywhere, under any $SU(N)/Z(N)$ irreducible representation. Implications of this in the 
	context of large $N$ reduction are discussed.
	\end{abstract}
	\date{\displaydate{date}}
	\maketitle

	\section{Introduction.}
	
	Pure $SU(N)$ lattice gauge theory with Wilson action on a periodic four torus is well defined by an integral over group elements. There is a local gauge invariance group and nothing special needs to be done about it. The integration includes all gauge copies. In numerical simulations one does not gauge fix because 
	locality is useful for importance sampling. Theoretically, one is free to partially or maximally fix a gauge. Only nonperturbative, exact, gauge fixing will be given consideration in this article. Perturbative BRST gauge fixing cannot be extended to exactly work on the lattice~\cite{brs2} in any simple way. 
	
	A transfer matrix $\mathfrak{T}=e^{-H}$ is needed in order to construct the finite temperature Gibbs ensemble. This paper is not concerned with $H$ itself because there is no continuous time. $\mathfrak{T}$ is constructed from the action in 
	a Euclidean path integral. For a purely bosonic system consisting of only site variables, finite temperature amounts to the imposition of periodic boundary conditions. 
	For gauge theories, $\mathfrak{T}$ is typically extracted from the path integral in a temporal gauge. At zero temperature
	one can set all time-like links to unity and $\mathfrak{T}$ is easily read off. At finite temperature the
	anticipated periodicity clashes with this gauge fixing because one cannot gauge away winding parallel transport.

	When read off directly from the path integral without fixing a gauge $\mathfrak{T}$~\footnote{Equation (3.144) in~\cite{MoMubook}, first paperback edition, 1997, Cambridge University Press, 1994.} does not come out in explicit form. One needs to integrate out the time 
	component of the gauge field exactly. For periodic boundary conditions one cannot avoid having to
	do this in order to get an explicit formula 
	for $\mathfrak{T}$.

   A deformation of the finite temperature periodic path integral was introduced in~\cite{UY} adding a term nonlocal in the euclidean evolution direction to the standard, local action. This complicates somewhat the extra integral needed to get $\mathfrak{T}$ explicitly. It will be shown that the deformation has a major effect on the Hilbert space $\mathfrak{T}$ acts in. 
	
   It was also claimed in~\cite{UY} that the deformation can be generalized to allow large $N$ Eguchi-Kawai~\cite{ek} reduction down to a single lattice site, all the way to continuum. Other proposals to achieve this end appear, superficially, to involve more drastic alterations. I shall argue that the modification of Hilbert space makes this newer proposal also drastic. 
	
    In order to make the proposals more comparable, the quenching prescription of~\cite{qek} will be adapted to provide a way alternative to that of~\cite{UY} to replace the canonical deconfined high temperature phase by a reconfined one.

	\section{Notation and generalities.}
	
	The Einstein summation convention won't be used and no meaning will be attached to whether an index is upper or lower. Summations are often implied by matrix products and trace operations. 
	
	Imagine a finite periodic four dimensional 
	hypercubic lattice embedded in and aligned with a continuum ${\cal T}^4$. 
	Its sites are labeled $x=(x_1 ,x_2 ,x_3 ,x_4)$ where $x_\mu=1,...,l_\mu$ are integer coordinates and the $l_\mu$ are lengths of sides in the $\mu$ direction. Directed links connect nearest neighboring sites from $x$ to $x+\mu$. In this context $\mu$ also denotes a unit vector in the $\mu$ positive direction. To each positively oriented link a variable $U_{x,\mu}$ is assigned. $U_{x,\mu}$ is an $N\times N$ unitary matrix of unit determinant. To a link from $x$ to negative $\mu$ direction we attach $U^\dagger_{x-\mu,\mu}\equiv U_{x,-\mu}$. When the site index $x$ is not present the reference will be to the collection of all links in the $\mu$ direction.
	Each link matrix is a faithful representation of one abstract element in $SU(N)$ and shall be
	identified with it. This is the fundamental representation of the group.
	To each site $x$ we attach an $N$-dimensional
	complex vector, $\phi_x$. The $\phi_x$ denote elements in a $C^N$ and collectively provide a space $\mathcal{V}$ on which matrices with $( N l_1 l_2 l_3 l_4 )^2$ entries act.
	These large matrices will be introduced later.
	The $\phi_x$ serve only this purpose and otherwise are ``unphysical''.  Here we deal only with gauge theory without site anchored matter. 
	
    The matrix entries $U_{x,\mu}^{i,j}, i,j=1,...N$ of $U_{x,\mu}$ 
    depend on $N^2-1$ real variables, but shall be viewed as a set of $N^2$ independent complex variables. The constraint to $SU(N)$ is implemented by a restrictions on the domain of integration over $U_{x,\mu}^{i,j}, i,j=1,...N$ which is embedded in $R^{N^2-1}$. The integration measure is left and right invariant Haar, and the group volume is set to unity. 
    
    One-link parallel transporters $T_\mu(U_\mu )$ act on $\mathcal{V}$ by
    \begin{equation}\label{eqn:def}
    [T_\mu(U_\mu)\phi]_x=U_{x-\mu,\mu}\phi_{x-\mu}
    \end{equation}
    The result is another vector 
    with entries defined by the equation employing implicit matrix-vector multiplication. 
    A $T_\mu$ has two indices, each of the form $(x,i)$ where $x$ is a site and $1\le i\le N$
    is a group index. The action of $T_\mu$ can be visualized as picking up the $N$ vector $\phi_x^j$
    and then parallel transporting it along the positively oriented link in the $\mu$-direction to 
    site $x+\mu$. Parallel transport makes the $N$ vector rotate in group space under the action of
    $U_{x,\mu}$. $T_\mu$ does this for all sites $x$. 
    Equation~(\ref{eqn:def}) is short-hand for
    \begin{equation}
    	[T_\mu(U_\mu)\phi]_{(x,i)}=\sum_y \delta_{y,x-\mu} \sum_{j=1}^N U_{y,\mu}^{ij}\phi_{(y , j)}
    \end{equation}
    $T_\mu$ can also be written as 
    \begin{equation}
    	T_\mu = T_\mu^0 \mathbf{U^\mu},
    	\end{equation}
    which separates group from site indices: $[\mathbf{U^\mu}\phi]_x^i =\sum_j U_{x,\mu}^{ij}\phi_x^j$.  $T_\mu^0$ generates cyclic rotation of sites 
    in the $\mu$ direction, $[T_\mu^0 \phi ]_x^i=\phi_{x-\mu}^i$. $T_\mu^0$'s for different $\mu$'s commute and can be simultaneously diagonalized. An action polynomial in norms of $[T_\mu,T_\nu]$ commutators 
    can be easily arranged to attain its minimum at $T_\mu=T_\mu^0$. In object oriented language 
    one can think of $T_\mu$ as a group-action overloaded cyclic shift $T_\mu^0$. 
    
    \subsection{Discrete Fourier Transform.}
    
    Consider a circle made out of $L$ sites, labeled by $q=0,...L-1$ in consecutive order in the positive direction. Let $v_q$ be the components of a site column vector $v$. The $T^0_\mu$ matrices with no group indices are replaced in this subsection by an $L\times L$ matrix $T^0$. 
    $T^0$ is diagonalized by the Discrete Fourier Transform (DFT) matrix $\Omega$:
    \begin{equation}
    	\Omega^{pq} = \frac{1}{\sqrt{L}} \omega^{-pq},\;\; p,q=0,1...,L-1 \;\;\; \omega =e^{-\frac{2\pi i}{L}}
    \end{equation}
    $\Omega$ is explicitly symmetric. The columns $p$ of $\Omega$ are orthonormal by:
    \begin{equation}
    	\frac{1}{L}\sum_p \omega^{qp}\omega^{-pq'}=\delta_{qq'}
    \end{equation}
    $\Omega$ is unitary, in addition to being symmetric. 
    The columns of $\Omega$ are eigenvectors of $T^0$:
    \begin{equation}
    \sum_q T^0_{q'q}\omega^{-pq}=\omega^p \omega^{-pq'}\;\;{\rm because}\;\sum_q T^0_{q'q} v_q = v_{q'-1}
    \end{equation}
    Hence	
    \begin{equation}
    	T^0 \Omega=\Omega \; {\rm diag}(1,\omega,..,\omega^{L-1}),\;\; \Omega^\dagger T^0\Omega = {\rm diag}(1,\omega,..,\omega^{L-1}),\;\; T^0 = \Omega \;{\rm diag}(1,\omega,..,\omega^{L-1})\Omega^\dagger
    \end{equation}
    $\Omega$ takes a site vector with components $v_q$ to ${\hat v}_p$, its DFT, and $\Omega^\dagger$ 
    inverts with $v =\Omega^\dagger {\hat v}$:
    \begin{equation}
    	(\Omega v )_p = {\hat v}_p =\frac{1}{\sqrt{L}} \sum_q \omega^{-pq} v_q ,\;\; (\Omega^\dagger {\hat v})_q=v_q=\frac{1}{\sqrt{L}} \sum_p \omega^{qp} {\hat v}_p
    \end{equation}
    Also,
    \begin{equation}
    [\Omega \;{\rm diag}(v_0, v_1,...,v_{L-1})\; \Omega^\dagger]_{pp'} =\frac{1}{L} \sum_q \omega^{-pq}  v_q\; \omega^{qp'}={\hat v}_{p-p'}
    \end{equation}
Taking the trace, one gets $\sum_q v_q=L{\hat v}_0$. 

When the $v_q$ are complex and independent, so are the ${\hat v}_p$; if the $v_q$ are real and independent, the complex ${\hat v}_p$ are constrained by ${\hat v}_p={\hat v}^*_{L-1-p}$. 

Arithmetic in the $p,q$ indices is modulo $L$ with the representative set $(0,1,...,L-1)$. All indices out of range are to be replaced by their representative in this set. Defining 
the $L\times L$ matrix $\hat V$ by $({\hat V})_{pp'}\equiv{\hat v}_{p-p'}$ shows that ${\hat V}$ is a circulant matrix, 
${\hat V} = {\rm circ}({\hat v}_0, {\hat v}_1,....{\hat v}_{L-1})$. Hence, conjugation by $\Omega ... \Omega^\dagger $ takes a diagonal matrix in site space to a circulant matrix in momentum space. Conjugation of $T^0$ by $\Omega^\dagger ... \Omega$ produces
a diagonal matrix in momentum space. The norm squared in matrix space of ${\hat V}$ is
\begin{equation}
	{\rm Tr} {\hat V}{\hat V}^\dagger = \sum_q |v_q|^2 = \sum_{p,p'} |{\hat v}_{p-p'}|^2=L\sum_p|{\hat v}_p|^2
	\end{equation}
In general, summation over $p$ and $p'$ 
of a summand that depends only on $r=p-p'$ gives a factor of $L$ times a summation over $r\equiv p-p'$  in the range $0$ to $L-1$. 

Circulant matrices are a natural way to ensure momentum conservation in diagrams. 
The indices $p$ and $p'$ provide a 't Hooft double lines notation, with Feynman propagator momenta given by $p-p'$. This is particularly useful in the case of planar diagrams, where planarity is a consequence of another index family, like color.

 On the full lattice introduced one employs a direct product of four DFT matrices $\Omega_\mu$
 each factor operating on an index $q_\mu=0,1,....,l_\mu-1$. The arithmetic in each $q_\mu$ is modulo $l_\mu$.

    \subsection{Gauge group.}
    
    To each site one also attaches a gauge transformation matrix $g_x\in SU(N)$. A collection of $g_x$ defines a gauge transformation on $\mathcal{V}$ given by:
    \begin{equation}
    [^{g}\mathfrak{G}\phi]_x=g_x\phi_x
    \end{equation}
    Then, 
    \begin{equation}
    [ (^{g}\mathfrak{G})^\dagger T(U_\mu) ( ^{g}\mathfrak{G} ) \phi ]_x =g_x^\dagger U_{x-\mu} g_{x-\mu}\phi_{x-\mu}
    \end{equation}
    
    We define the action of a gauge transformation on the link variables as the 
    replacement of $U_{x,\mu}$ by $^{g}U_{x,\mu} $:
    \begin{equation}
    ^{g}U_{x,\mu}= g_{x+\mu}^\dagger U_{x,\mu}  g_x   
    \end{equation}
    Then, the action of gauge transformations on parallel transporters is by conjugation: 
    \begin{equation}
    (^{g}\mathfrak{G})^\dagger (T_\mu (U_\mu) ^{g}\mathfrak{G} = T_\mu (^{g}U_\mu)
    \end{equation}
    The matrices $^{g}\mathfrak{G}$ make up the gauge group:
    \begin{equation}
    ^{gh}\mathfrak{G}={^{g}\mathfrak{G}} {^{h}\mathfrak{G}}, 
    (^{g}\mathfrak{G})^\dagger={^{g^{\dagger}}\mathfrak{G}}=({^{g}\mathfrak{G}})^{-1}
    \end{equation}
    The norm of some $X$ in the algebra generated by the $T_\mu$, $||X||^2=Tr X X^\dagger$, is gauge invariant. 
    
    \subsection{Action in path integral.}
    
    The Wilson partition function is the simplest possible on a hypercubic lattice:
    \begin{equation}
    \label{eqn:Z}
    Z_W=\int [\prod_{x,\mu} dU_{x,\mu}] e^{-\frac{\beta}{2} \sum_{\mu < \nu} Tr[ [T_\mu (U_\mu),T_\nu(U_\nu)][T_\mu(U_\mu),T_\nu(U_\nu)]^\dagger]}
    \end{equation}
    $\beta$ is the lattice coupling constant.  The trace is a
    sum over site and group indices. Before the trace is taken the expressions (dropping the arguments) read:
    \begin{equation}\label{eqn:comm}
    [T_\mu ,T_\nu][T_\mu ,T_\nu ]^\dagger = 2
    - T_\mu T_\nu T_\mu^\dagger T_\nu^\dagger - T_\nu T_\mu T_\nu^\dagger T_\mu^\dagger = (1-T_\mu T_\nu T_\mu^\dagger T_\nu^\dagger)+[{\rm h.c.}]
    \end{equation}
    
    The term proportional to the unit matrix can be dropped from the action. The remaining 
    two terms represent parallel transports round local small closed square circuits in the $\mu-\nu$ plane, diagonal in 
    lattice sites, circulating in opposite directions in the $\mu-\nu$ plane. 
    In general, a circuit which does not close gives a parallel transport matrix with zeros on the diagonal because the site components of the indices are unequal. Eight distinct choices for the 
    first commutator (which then determines the second) are possible after choosing one $\mu$--$\nu$ plane (if $\mu=\nu$ the commutator vanishes and there is no plane.)
    The first term has two possible choices for direction and independently two choices for sense.
    The second term has a direction fixed by the first and two choices for sense. The direction is $\mu$ 
    or $\nu$ and the sense is positive (no dagger) or negative (with dagger). Everyone of the eight choices
    produces the same action. These eight possibilities split into two groups of four, according to the sign of
    circulation in the $\mu$--$\nu$ plane.

    \subsection{Connection to continuum.}
    
    We assume a continuum $su(N)$ matrix valued 
    one form $A=A_\mu(\mathpzc{x})d\mathpzc{x}_\mu$ is given and define a continuum $\phi(\mathpzc{x})$
    on the ${\cal T}^4$ torus the lattice is embedded in. The spacing between nearest neighboring lattice points  
    is uniformly equal to $a$. Let $D_\mu (A_\mu)$ be the covariant derivative  associated with $A$ in 
    the direction $\mu$. It acts on fields $\phi(\mathpzc{x})$. Consider the action of $e^{aD_\mu (A_\mu )}$ on 
    $\phi(\mathpzc{x})$. Observe that $[e^{aD_\mu (A)}] (\phi) (\mathpzc{x})$ restricts to $\phi_x$ with $x=\mathpzc{x}$: if we know the continuum function $\phi(\mathpzc{x})$ 
    at the sites $\mathpzc{x}=x$ the action of $e^{aD_\mu (A_\mu )}$ on $\phi(\mathpzc{x})$
    will produce values again localized on sites. The in-between points do not enter directly. The action
    of $e^{aD_\mu (A_\mu )}$ restricts to sites where it becomes the action of $T_\mu (A_\mu )$.
    With this restriction we have
    \begin{equation}
    T_\mu (A_\mu ) = e^{aD_\mu(A_\mu)}
    \end{equation}
    Consequently, formally,  $[T_\mu (U_\mu),T_\nu(U_\nu)] = a^2 [D_\mu (A_\mu), D_\nu (A_\nu)]+... \equiv a^2 F_{\mu\nu} (A) + ...$, establishing the formal connection of the Wilson action to the continuum one. Note that the  commutator $[T_\mu ,T_\nu ]$ 
    is not simply related to its hermitian conjugate, nor is it site diagonal. Thus, in itself, it is not a good choice 
    for representing the local non-abelian field strength on the lattice. Combining terms coming from the
    alternative forms mentioned after equation~(\ref{eqn:comm}) provides a set of eight circuits which have just one site index shared by all eight and keeping them untraced, one generates eight objects transforming under the adjoint representation of the gauge group at the common site. Subtracting the sum of four 
    terms with one circulation from the sum of terms of opposite circulation provides 
    a good lattice version of the continuum nonabelian field strength. It is known as the SW~\cite{sw} (or clover) term.

    \subsection{Large $N$ reduction.}
    
    Gauge transformations act by the same conjugation simultaneously on all four $T_\mu$'s. Any function of the 
    link variables that is expressible as the trace of a product of $T_\mu$'s and their
    conjugates is gauge invariant. These observables are summed over all lattice sites.
    Their average expectation values are obtained after division by $\prod_\mu l_\mu$. 
    When the number of $T_\mu$'s minus that of $T_\mu^\dagger$'s does not
    vanish modulo $l_\mu$ for all $\mu$ every diagonal entry of the product matrix vanishes  
    and the trace is zero. 
    
    Eguchi-Kawai~\cite{ek} reduction is the statement that in the large $N$ limit of the theory the expectation values of these variables are identical to those of the theory with $l_\mu=1$ for all $\mu$. With a single available site, the $T_\mu$'s become
    just complete unitary matrices. If the number of sites is much larger than $N$ the $T_\mu$ matrices
    are sparse. In the regime that Eguchi-Kawai reduction holds this spareness becomes irrelevant
    and only a trivial dependence on the volume remains. This is vaguely reminiscent of the eigenvalue distributions of single random banded Hermitian matrices~\cite{bm}. 
    
    In quenched Eguchi-Kawai reduction~\cite{qek}, where there is only one site, $T_\mu = U_\mu$ transforms in the adjoint under the gauge group, so its eigenvalues are 
    gauge invariant. The minimum of the action is attained when all $[ U_\mu , U_\nu ]=0$
    which means they are simultaneously diagonalizable. This allows to separate out the
    eigenvalues and by quenching them they are capable of playing the role
    of lattice momenta in perturbation theory. 
    
    In the clever model of twisted Eguchi-Kawai reduction~\cite{twist}, the
    action minimum is arranged to be attained at $\theta_{\mu\nu}$-twisted configurations -- $U_\mu U_\nu =e^{i\theta_{\mu\nu}} U_\nu U_\mu$ -- where a choice of twists $e^{i\theta_{\mu\nu}}=e^{-i\theta_{\nu\mu}}$ is made from the start. Solutions to 
    these relations produce a system on noncommutative Euclidean space in the continuum limit. 
    In perturbation theory this is detectable only at sub-leading order 
    in $1/N$, because in planar-diagram the twists cancel out and the untwisted series is recovered. 
    Quenched and twisted models are drastic alternatives to~\cite{ek} alluded to earlier.

     \section{Transfer matrix and Hilbert space.}
     
     In a quantization that starts from a path integral one foliates 
     the space of configurations into co-dimension one leaves of identical
     structure which are connected in the action only in sequential pairs. 
     The (Euclidean) evolution takes one leaf to the next one. The kernel is 
     expressed in terms of the same variables as the path integral. Thus, the 
     Hilbert space is defined as the space of integrable functions of the variables 
     resident on one leaf. For lattice gauge theory the functions are on a compact 
     space -- there is a discrete countable basis. The inner product of the Hilbert space
     comes from the ${\cal{L}}_2$ norm on complex functions on the space-link space. 
     
     On our hypercubic lattice we pick the evolution direction as $\mu=4$.
     The space-leaves are three dimensional tori of sizes $l_1\times l_2\times l_3$ 
     at fixed $t\equiv x_4$. The kernel 
     is homogeneous along $t$ with its two arguments determined by variables 
     appearing in one $(t,t+1)$ pair. 
     The $t$ variables is shared with the preceding kernel 
     and the $t+1$ ones with the following one. Integration over shared variables is 
     factorized by the completeness identity of the countable basis in which $\mathfrak{T}$
     has discrete indices. 
     This basis provides discretely valued indices for the transfer matrix. Sometimes 
     a more formal labeling by continuous variables is used. 
     Integration in the path integral is ordered ascending in $t$. Periodic boundary
     conditions close the circle cyclically. 
     The path integral becomes a trace of the transfer matrix $\mathfrak{T}$ raised to integral power. 
      
     For a theory with only site anchored variables and nearest neighbor
     interactions one can read $\mathfrak{T}$ off the path integral and the Hilbert
     space is determined. 
     In gauge theory, one needs to first integrate out all link variables
     connecting adjacent leaves. Only then can $\mathfrak{T}$ be extracted explicitly; this will be shown in great detail.

     For a general action this integral may be complicated. In gauge theory,
     there is gauge invariance under transformations parametrized
     by site variables in four space.  We only consider gauge invariant observables, 
     and we could just integrate over link variables modulo gauge transformations. 
     The number of intra leaf link matrices is equal to that of gauge variables. 
     As expected, the integration over leave--connecting links giving $Z_W$ (eq. \ref{eqn:Z}) is doable and  restricts to a gauge invariant ecosystem. 
     
     With $\mathfrak{T}$ in hand, one can deal with excitations by inserting operators
     in the trace. Some observables may contain time-link variables, 
     which were eliminated when $\mathfrak{T}$ was constructed. In order to reproduce answers for all expectation values
     accessible to the path integral formalism one must revisit 
     the integration over link variables appearing in observables.
     
     \subsection{Explicit construction}
 
      The separation between time-link and space-link variables requires some new notation.
      The positive space directions emanating from site ${\vec x}=(x_1, x_2 , x_3 )$ on one leaf 
      will be denoted by $\alpha,\beta ...$. 
      The distinct space leaves will be labeled by $t$. Space link variables on leaf $t$ will be denoted $U^t_{{\vec x},\alpha}$; when the entire set of a leaf is intended, the
      index $\vec x$ will be dropped.\footnote{The notation $X^t$ for the transpose 
      	of a matrix $X$ shall not be employed in this paper: superscript $t$ 
      	refers to the leaf at ``time'' $t$.} The action of $T^t_\alpha$ is restricted to leaf $t$; they  are
      inter leaf operators. The link-matrix arguments of the in-leaf parallel transporters $T^t_\alpha (U^t_\alpha)$ 
      will be dropped.  
      Three dimensional space gauge transformations are similarly  restricted to act
      only on leaf $t$. They shall be denoted by $^{g}{\mathfrak{G}}_t$ and act by 
      $( ^{g}{\mathfrak{G}}_t )^\dagger T_\alpha (U^t_\alpha) ^{g}{\mathfrak{G}}_t = T^t_\alpha (^{g}U^t_\alpha)$.

      The action term $S$ in the exponent of (\ref{eqn:Z}) is split into an inter leaf and
      an intra leaf part:
      
      \begin{align}    
      	\begin{split}  	
      	S=&\sum_{\mu<\nu}Tr{\big [} [T_\mu ,T_\nu ][T_\mu,T_\nu ]^\dagger{\big ]} = \\ \sum_t 
      	&\{\sum_{\alpha<\beta} Tr{\big [} [T^t_\alpha ,T^t_\beta][T^t_\alpha,T^t_\beta]^\dagger{\big ]}+
      \sum_\alpha Tr{\big [} [T_{t+1/2} ,T^t_\alpha][T_{t+1/2},T^t_\alpha]^\dagger{\big]}\}
      \end{split}
      \end{align}
      
      $T_{t+1/2}$ is the parallel transporter from leaf $t$ to $t+1$ in the $\tau$ direction.
      Transport in the opposite direction is $(T_{t+1/2})^\dagger$. 
      The (suppressed) argument of $T_{t+1/2}$ is the collection over $\vec x$ in leaf $t$ of $U^t_{\vec x , t}$ 
      and that of $(T_{t+1/2})^\dagger$ is the collection over $\vec x$ in leaf $t+1$ of $U^{t+1}_{\vec x , t}$.
      It is convenient to replace $U^t_{\vec x , t}$ by $g_{t+1/2} ({\vec x} )$ 
      because one can view these matrices 
      as generating an inter leaf $t+1/2$ gauge transformation by $U^{t+1}_{{\vec x},\alpha} \rightarrow [g_{t+1/2}({\vec x })] U^{t+1}_{{\vec x}, \alpha} [g_{t+1/2,\alpha} ({{\vec x}+{\hat{\alpha}}})]^\dagger $ 
      or, equivalently, $U^{t}_{{\vec x}, \alpha} \rightarrow [g_{t+1/2}({\vec x })]^\dagger
      U^{t}_{\vec x, \alpha} g_{t+1/2}({\vec x} +\hat{\alpha}) $.
      We shall use $T^0_{t+1/2}$ to denote the gauge field independent lattice shift forwards. Its dagger is the  shift backwards. 
      
      $S=\sum_t S(t,t+1)$ where each term is associated with the pair ($t$,$t+1$). $S(t,t+1)$ 
      is made out of two inter leaf terms, $S(t)$, $S(t+1)$ and one intra leaf term, $S(t+1/2)$.
      \begin{equation}
      S(t)=\frac{1}{2} 	\sum_{\alpha<\beta} Tr {\big [} [T^t_\alpha ,T^t_\beta][T^t_\alpha,T^t_\beta]^\dagger{\big ]}
      \end{equation}
      \begin{equation}
      S(t+1/2)=\sum_\alpha Tr{\big [} [T^0_{t+1/2} ,T^t_\alpha(U^t_\alpha) ]
      [T^0_{t+1/2},T^t_\alpha( ^{g_{t+1/2}}U^{t+1}_\alpha)]^\dagger]{\big ]}
      \end{equation}
      
      $S(t)$ is a $^{g}{\mathfrak{G}}_t$ invariant function $\mathfrak{S}_1 (U^t_{{\vec x}, \alpha})$.
      $S(t+1/2)$ is a function 
      ${\mathfrak{S}}_2 ( U^t_{{\vec x},\alpha}, U^{t+1}_{{\vec x}, \alpha}, g_{t+1/2}({\vec x}))$ 
      which can be viewed as a function 
      ${\mathfrak{F}}^A_2 (^{g_{t+1/2}({\vec x})}{U^{t+1}_{{\vec x}, \alpha}}, 
      	U^t_{{\vec x},\alpha})$ or as function ${\mathfrak{F}}^B_2 
      	(^{[g_{t+1/2}({\vec x})]^\dagger}	U^t_{{\vec x}, \alpha},U^{t+1}_{{\vec x}, \alpha})$.

      The partition function can be rewritten as:
      \begin{equation}      \label{eqn:X}
      \begin{split}
      Z_W= &\int [dg^1]....[dg^{l_4}]\int [dU^1]....[dU^{l_4}]\; e^{-\frac{\beta}{4} 
      	[\mathfrak{S}_1 (U^1_{\vec x, \alpha})+\mathfrak{S}_1 (U^2_{\vec x, \alpha})]}
       e^{-\frac{\beta}{4} [\mathfrak{S_2}(U^1_{\vec x,\alpha}, U^2_{\vec x, \alpha}, g^{2}_{\vec x})]}\\
      &....e^{-\frac{\beta}{4} 
      	[\mathfrak{S}_1 (U^t_{\vec x, \alpha})+\mathfrak{S}_1 (U^{t+1}_{\vec x, \alpha})]}
       e^{-\frac{\beta}{4} [\mathfrak{S_2}(U^t_{\vec x,\alpha}, U^{t+1}_{\vec x, \alpha}, g^{t+1}_{\vec x})]}
       e^{-\frac{\beta}{4}
      	[\mathfrak{S}_1 (U^{t+1}_{\vec x, \alpha})+\mathfrak{S}_1 (U^{t+2}_{\vec x, \alpha})]}     
      e^{-\frac{\beta}{4} [\mathfrak{S_2}(U^{t+1}_{\vec x,\alpha}, U^{t+2}_{\vec x, \alpha}, g^{t+2}_{\vec x})]}\\
      &....e^{-\frac{\beta}{4}
      	[\mathfrak{S}_1 (U^{l_4-1}_{\vec x, \alpha})+\mathfrak{S}_1 (U^{l_4}_{\vec x, \alpha})]}      
      e^{-\frac{\beta}{4} [\mathfrak{S_2}(U^{l_4-1}_{\vec x,\alpha}, U^{l_4}_{\vec x, \alpha}, g^{l_4}_{\vec x})]} e^{-\frac{\beta}{4}
      	[\mathfrak{S}_1 (U^{l_4}_{\vec x, \alpha})+\mathfrak{S}_1 (U^1_{\vec x, \alpha})]}      
      e^{-\frac{\beta}{4} [\mathfrak{S_2}(U^{l_4}_{\vec x,\alpha}, U^{1}_{\vec x, \alpha}, g^{1}_{\vec x})]}
      \end{split}
      \end{equation}
      The $U$ integrals are manipulated first, at fixed $g$. We make a change of integration 
      variables $U^t_{\alpha}\rightarrow ^{g_{t+1/2}}\!U^{t}_\alpha$. This does not affect the $\mathfrak{S}_1$ terms. In the $\mathfrak{S}_2$ terms it replaces $g^t_{\vec x}$ by ${\mathbf 1}$. One has to keep 
      the $^{g_{t+1/2}}U^{t}_\alpha$'s in the integration measure because the  $\mathfrak{S}_2$ are not
      invariant under the change of integration variables. This is a round about way of fixing 
      a gauge in the four dimensional sense, however, it holds also on the torus, despite the fact that
      parallel transport along a winding curve cannot be gauged away. The  
      integration over the $g$ variables has been moved  from the kernel to the measure. 
      
      \subsection{Gauge averaging per leaf.}

      One now factorizes the integrals over link 
      variables with the completeness relation shared between consecutive pairs of leaves and ends up with a product over infinitely dimensional matrices replacing equation (\ref{eqn:X}).
      Periodicity implies that a trace is taken. 
      
      For pure $SU(N)$ gauge theory the space is a tensor product over one Hilbert space for each matrix variable, $V$, consisting of complex valued functions of $V$:
      \begin{equation}
      (\Psi,\Phi)=\int[dV] \overline{\Psi(V)}\Phi(V)
      \end{equation}
      The single link Hilbert space has a basis consisting of multivariate 
      polynomials of increasing degrees in the entries of $V$ and their complex conjugates. 
      Elements of the
      basis are labeled by a composite index $\mathfrak{R}= (R, i, j)$ where $R$ denotes 
      a unique irreducible representation of dimension $d_R$ and $1\le i,j \le d_R$ label the 
      entries: $\Psi_\mathfrak{R} (V) =\sqrt{d_R}[{\cal D}_R (V)]_{ij}$. The polynomials 
      ${\cal D}_R (V)_{ij}$ are determined by the associated Young diagram. The completeness relation is 
      \begin{equation}\label{eqn:compact}
      \sum_\mathfrak{R} \Psi_\mathfrak{R} (V) \overline{\Psi_\mathfrak{R} (U)} =\delta_{SU(N)}(V,U)=\delta(VU^\dagger,\mathbf{1}),
      \end{equation}
      where the delta-function is on group elements. Hence the full Hilbert space is spanned 
      by a basis of multinomials, consisting of products of representation polynomials, one for each $R$ per link. $U$ and $V$ now denote collections over all link variables 
      in a leaf. On the left hand side the sum over $\mathfrak{R}$ is a sum over all possible combinations of ${\mathfrak{R}}_{\vec{x},\alpha}$, one for each $(\vec{x},\alpha )$ link on a leaf. 
      Similarly, $\Psi_\mathfrak{R} (V)$ and $\Psi_\mathfrak{R} (U)$ each are a product over such factors, one for each link. Finally, he right hand side of the equation is a product of
      one $\delta_{SU(N)}$ for each link on the leaf.  Equation~(\ref{eqn:compact}) replaces each
      integrations over shared space-like link variables by a factorized form and the path becomes 
      a matrix product of identical factors $\mathfrak{T}$.

      I now turn to the more economical Dirac notation with states $\ket{U}$ where $U$ denotes the entire inter link configuration of a space leaf. Then the generalized $\delta_{SU(N)}(V,U)$ is $\braket{V|U}$ 
      with $\bra{V}$ in the dual space to $\ket{U}$ and the decomposition of unity is $\int [dU] \ket{U}\!\bra{U}$.

      Our task is to evaluate the integration over the $g$ variables at fixed $U$'s.
      The dependence on $g$'s resides only in the measure. Consider $\mathfrak{P}$, an operator given below:
      \begin{equation}
      \mathfrak{P}=\int [dg] \sum_\mathfrak{R} \Psi_\mathfrak{R} (V) \overline{\Psi_\mathfrak{R} (^{g}U)}
      \end{equation}
      The integration variables consist of one $g_{\vec x}$ for each site on one leaf. Acting with 
      $\mathfrak{P}$ on a state $\Psi (^{g}U)$ we obtain a state $\Phi (U)$:
      \begin{equation}
      \Phi(V)=(\mathfrak{P} \Psi )(V)= \int [dg] \int [d ^{g}\!U]\braket{V | ^{g}\!U} \Psi(U)= 
      \int [dg] \Psi (^{g^\dagger}\!V)=\int [dg] \Psi (^{g} V)
      \end{equation}
      Obviously, $(\mathfrak{P} \Phi)(V)= \Phi (V)$, so $\mathfrak{P}^2=\mathfrak{P}$. The operator $\mathfrak{P}$ projects on the gauge singlet space of the leaf gauge group. The same projector 
      can be represented by using the tangent vector fields associated with infinitesimal 
      space-leaf gauge transformations. This would give the Gauss law constraints on states 
      appearing in the Kogut-Susskind (KS) continuous time Hamiltonian defined on a spatial lattice. The KS 
      construction was built from scratch, without relying directly on the Wilson path integral. The intra leaf term is now given by a group manifold Laplacian for each leaf--link variable. It is constructed from the same vector fields entering Gauss law constraints. Gauss' law has nothing to do with continuous, as opposed to discrete, time.
      
      We could first introduce an unrestricted transfer matrix $\mathfrak{T}_0$ starting from the exponentiated Kogut-Susskind Hamiltonian and forgetting about Gauss' law. 
      The restricted $\mathfrak{T}$ we get from the requirement to reproduce
      exactly the path integral is then  $\mathfrak{P}\; \mathfrak{T}_0 \;\mathfrak{P}$. $\mathfrak{T}_0$ is given by:
      \begin{equation}
      \begin{split}
      \bra{V}\mathfrak{T}_0\ket{U}=& e^{-\frac{\beta}{4}\sum_{\alpha < \beta} Tr[ [T_\alpha (V_\alpha ),T_\beta (V_\beta )][T_\alpha (V_\alpha ),T_\beta (V_\beta)]^\dagger ]}\\
 &e^{-frac{N}{8\lambda}\sum_{\alpha<\beta} Tr[ [T_\alpha (U_\alpha ),T_\beta (U_\beta )][T_\alpha (U_\alpha ),T_\beta (U_\beta)]^\dagger ] }\\      
      &e^{-\frac{\beta}{2} \sum_\alpha
      	Tr[[ T_\alpha (V_\alpha ),T_t^0 ][T_\alpha (U_\alpha ),T_t^0 ]^\dagger]]}\\   
      &e^{-\frac{\beta}{4}\sum_{\alpha<\beta} Tr[ [T_\alpha (U_\alpha ),T_\beta (U_\beta )][T_\alpha (U_\alpha ),T_\beta (U_\beta)]^\dagger ] }=\mathfrak{T}_0(V,U)=\mathfrak{T}_0(U^\dagger,V^\dagger)
      \end{split}
      \end{equation}    
      In terms of $\mathfrak{T}$, $Z_W$ is given by $Tr \mathfrak{T}^{l_4}$.

      Alternatively, start with an $l_4=\infty$ system, fix the gauge $U_{\vec{x},t}=
      \mathbf{1}$ and adopt $\mathfrak{T}_0$ from there. Next, prove that 
      $[\mathfrak{T}_0 ,\; ^{g}G_t]=\mathbf{0}$ for any $t$. For finite $l_4$, with periodic boundary
      conditions, one needs to keep a nontrivial $U_{\vec{x},t}$ at one $t$, say at $t=l_4$. 
      Integrating over the nontrivial $U_{\vec{x},l_4}$ forces the insertion 
      $\mathfrak{P}$ only at $t=l_4$. $\mathfrak{P}$ can then be 
      propagated to all other locations along the $t$ span by using ${\mathfrak{P}}^n =\mathfrak{P}$ for any positive integer $n$ and $[\mathfrak{P},\;{\mathfrak{T}}_0]=0$. 
      So, fixing the gauge correctly produces the same result as not picking any gauge at all, as it should be.

      \subsection{Observables containing time-like link matrices.}
      
      The obvious observable set we have access to consists of multinomial functions of link variables that are invariant under $\mathfrak{P}$. These are traces of parallel transport round closed loops, both contractible and not. 
      Let $\Op$  be such an  observable on one leaf. Then,
     \begin{equation}\label{eqn:ginvobs}
     	{\braket{\Op}}_{path-integral}= \Tr (\mathfrak{T} \Op ).
     	\end{equation}
     
      So far, the set of observables does not include Wilson loops with time-like links. 
      Time like links in a gauge invariant observable are fine in the path integral. Hence, observables containing time like links must be allowed also in transfer matrix language. The presence of time like links in an observable affects the integration
      over time like link matrices. The latter were absorbed into three dimensional gauge transformations acting on individual links appearing in the action. Equation~\ref{eqn:ginvobs} does not hold anymore. The time links
      present in the observable would be washed out by specific factors in $\mathfrak{P}$. 
       
       This easy to correct. Let us take the simple example of a localized rectangular Wilson loop in the $x_1 - x_4$ plane with an extent $l$ in the $x_1$ direction and $t$ in the $x_4$ one. In this case one no longer 
       averages over all space anchored gauge transformations. The factors in $\mathfrak{P}$
       acting at the spatial locations of the corners separated only in the $x_4$ direction are replaced by projectors on 
       polynomials of space link variables which transform as ${\mathfrak R}=F$ and its
       conjugate respectively. Here, $F$ denotes the fundamental representation. To work this
       out explicitly one uses the basis of equation~\ref{eqn:compact}. One can then generalize
       and replace the two lines in each leaf by any contour connecting the same endpoints. If the separation $t$ is large enough (but still smaller than $l_4/2$ ) the dominating state will contain a 
       superposition of many paths connecting the corners and arbitrary extra closed contours. For large enough $t$ the state of lowest energy is in the sector of Hilbert 
       space selected by the amended projectors will dominate. 
       This state, with fixed ends and in the limit of large separations is an object effective open string theory could describe approximately.

       There is a qualitative difference between observables and action: while both are 
       functions of space links, the observables consists of a relatively small number of local ``disturbances'' over the vacuum, which is the ground state of $\mathfrak{T}$. 
       More precisely, when the spatial volume is taken into infinity, the observable depends
       on only a finite number of spatial links. This is analogous to particle excitations above the vacuum in non-gauge theories.

      \subsection{Finite temperature deconfinement transition.}
      
      Consider again the open string state described in the previous section. 
      Its energy relative to the sourceless vacuum state grows linearly with the separation $l$ at large enough $l$. This is an asymptotic statement: one would
      need to take the space directions to infinity for a precise formulation. This will 
      hold at low enough temperatures, but ceases to be true above a transition temperature, 
      where the theory deconfines. The presence of such a transition for stringy states has been known before lattice gauge theory was invented~\cite{hagedorn}. The point is that the number of different 
      contributions (orthogonal states entering additively in the trace) increases exponentially with $l$ because a meandering curve anchored at 2 widely
      separated states can change its shape anywhere locally along it while still obeying the constraint to fixed end points. The entropy can overcome the suppression coming from the energy at high enough temperature. When this happens the string ``dissolves''. 
      
      The clearest symptom of this has to do with an extra global symmetry. For zero temperature this
      symmetry is part of the gauge group; for non-zero temperature it is separate. 
      Every spatial location defines a minimal length parallel transporter round the periodic
      direction which can be multiplied by a space independent $N^{\rm th}$-root of unity, $z$, without changing the action. In equations, $T_4 \rightarrow zT_4$, with $z\in Z(N)\subset SU(N)$. 
      This is a global (in space) $Z(N)$ symmetry. The string state has sources in conjugate
      fundamentals at its distinct ends, so also remains invariant. After
      deconfinement no correlation should be left between the two sources at the string ends when 
      they are widely separated. The expectation value of the observable should cluster into a product 
      of two expectation values, each associated with one of the strings ends. Each one of these factors no longer is invariant under the $Z(N)$. So, unless the correlation is exactly zero, the global $Z(N)$ is
      spontaneously broken at infinite spatial volume. 
      
      The symmetry breaking feature of the finite temperature transition has led
      to an industry of effective $Z(N)$ models whose variables could be thought of 
      as stand-ins for traces of winding (Polyakov-) loops. Approximate renormalization
      group methods~\cite{ogil} indicated that such effective models have some validity and this physical picture has accumulated support over the years and is widely accepted today. 
      
      One can also consider an observable consisting of two parallel minimal length thermal loops separated in space and winding once in opposite directions. Now, 
      the partition function in the presence of sources depends on their separation and the energy--entropy argument applies again. $Z(N)$ breaking by clustering is directly evident.
       
      \subsection{Eguchi Kawai reduction and its connection to finite temperature.}
      
      Eguchi Kawai reduction has already been introduced. The four dimensional Euclidean system 
      consists of just four $SU(N)$ unitary matrices associated with four loops, all 
      beginning and ending at the same single point. Each loop has an associated 
      $Z(N)$ symmetry. The original Eguchi Kawai claim was that this model exactly reproduces 
      infinite $N$ and arbitrary volume $SU(N)$ lattice gauge theory with Wilson 
      action. For their argument to hold, preservation of the $Z(N)$ symmetry at is crucial. It turned out that the $Z(N)$'s will break also as $N\to\infty$, once the physical gauge coupling is weak enough.
      
      It was understood that the role of momentum space might be taken over by the
      eigenvalue sets of the four link matrices, but that would work only if these 
      eigenvalue sets were individually and independently uniformly distributed, which, 
      would ensure the preservation of the four $Z(N)$'s also at infinite $N$. The failure of this to happen could be understood from a one loop calculation of the effective potential
      governing the dynamics of these eigenvalues. In the quenched version this was
      handled by redefining the model requiring the eigenvalues to be quenched and thus
      neutralizing the induced effective potential. For small $\beta$ 
      the $Z(N)$'s did not break at infinite $N$ and quenching had no effect. Hence, the validity of the original Eguchi Kawai analysis was 
      preserved in the strong coupling regime. The extension to weak coupling by quenching 
      reproduced normal infinite volume lattice perturbation theory and was conjectured to
      ``fix'' EK-reduction. The amended model is called the ``quenched Eguchi-Kawai'' (QEK)  model. 
      
      While the finite temperature transition occurs only at infinite spatial volume 
      with any finite $N$, the one here takes place, because $N$ was taken to infinity, even at finite spatial volume.
      Nevertheless, the explicit calculation of the effective potential made it obvious 
      that the mechanisms were similar. This was shown in~\cite{fintempnpb}. I present this derivation below in a somewhat more general case 
      than in the original paper, because it is related to 
      ``deconstruction''~\cite{deconstruction}. Also, it is more detailed, emphasizing the mechanism by which taking $N\to\infty$ first eliminates volume dependence.

      \subsection{Summary of section.}
      
      The restriction to space-leaf gauge invariant states in the transfer matrix formalism 
      is a direct consequence of periodicity in the path integral. If the time extent is 
      infinite one can use a partially fixed gauge $U_{{\vec x}, t}=\mathbf{1}$ and read off the
      transfer matrix with ease. For a finite periodic extent the same transfer matrix works but
      now the projector $\mathfrak{P}$ must be introduced in the trace. This can be obtained by
      now fixing all $U_{{\vec x}, t}=\mathbf{1}$ but one, and integrating over the unfixed link variable.
      This framework extends to the inclusion of all observables accessible to the path
      integral by exploring local alterations of the projectors $\mathfrak{P}$.
      
      The transfer matrix formalism leads naturally to a physical picture predicting 
      deconfinement above some finite temperature. The transition reflects itself in 
      spontaneous breaking of a global $Z(N)$ symmetry in the path integral formalism.
      
      Eguchi-Kawai reduction has all four directions compact and therefore four independent
      $Z(N)$'s. They need to all be preserved for it to work. It turns out that 
      these symmetries do break spontaneously at large $N$, even though the four volume
      is minimal. It was later found numerically that if all four directions are large enough in physical, continuum units (but far from absurdly large) the desired symmetry remains preserved in the continuum limit~\cite{contek}. So, Eguchi-Kawai volume independence is likely valid in continuum, but only for volumes whose sides are large enough. 

\section{Finite temperature at weak gauge coupling.}

In this section we slightly change notation and focus: We single out one direction whose periodicity is $N_t$, the lattice inverse temperature. In the spatial directions all periodicities are $L$. The spatial lattice has $d$ dimensions; of special interest are the ultraviolet complete cases $d\le 4$. 

\subsection{Complete gauge fixing at finite temperature.}

This time, the gauge will be completely fixed. All directions in the space of remaining 
variables will have bounded fluctuations in the Gaussian approximation, eliminating the need 
for ghosts, without abandoning validity outside perturbation theory, where BRST 
gauge fixing does not apply~\cite{brs2}.

Consider a ring of $N_t$ sites connected by links on which $SU(N)$ unitary matrices
$g_0, g_1,...,g_{N_t-1}$ reside. This ring is perpendicular to each leaf at the site located 
at ${\vec x}$ on that leaf. Start at site $2$ and make a gauge transformation there by $g_1$.
$g_1$ acts backwards on the link $1-2$ and replaces $g_1$ there by $\mathbf{1}$. It also acts 
on the leaf, but the action does not change and neither does the Haar integration measure. It acts forward on link $2-3$, replacing $g_2$ that resided there by $g_1 g_2$. Now we make a gauge 
transformation by $g_1 g_2$ at site $3$ which puts unity on link $2-3$ amd $g_1 g_2 g_3$ on 
link $3-4$. When we reach site $N_t-1$ this process is done one last time, with the product
$g_1 g_2 ....g_{N_t-1}$ being placed on link $N_t-1$. We cannot go further, because this would affect link 
$0$. This is as it should be because the product $g_1 g_2 ....g_{N_t-}$ transforms by conjugation 
under the gauge group, so its set of eigenvalues is gauge invariant: a multiplication from only one side would change these. It is still possible though to do a gauge transformation by an $SU(N)$ matrix $W$ at all sites on the ring. This leaves all the links
we set to unity unchanged because they get a $W$ from one end and a $W^\dagger$ from the other.
On link $(N_t -1)-0$ the variable $g_0 g_1 ....g_{N_t -1}$ gets conjugated. We can choose $W$ to be the diagonalizing matrix. We fix $W$ up to a discrete element of $Z(N)$ to produce a diagonal matrix
on link $(N_t -1)-0$ whose angles increase from $-\pi$ to $\pi$ as their index increases from $0$ to $N_t-1$. 
The remaining $Z(N)$ can be fixed by forcing one entry of $W$ into a chosen interval of 
length $\frac{2\pi}{N}$ on the unit circle. The change of variables from $g\equiv g_0 g_2....g_{N_t-1}$
to $W$ and $\theta_j$ that takes place at each $x$ is accompanied by the usual Jacobian. 
There is one such factor for each spatial site and it has preference for an equally spaced distribution
of the $\theta_j$'s round the unit circle. One can move all $W$'s along the unit links until they 
completely disappear. Only the Jacobian factors, one for each $x$, remain.

In perturbation theory we want translational invariance in all directions because then we can 
simultaneously diagonalize all $T_\mu$'s at the expansion point using the DFT matrices, as discussed. To achieve translational invariance in the $\tau$ direction we allow 
the introduction of diagonal unitary link matrices not in $SU(N)$: their determinant is
not unity. There is no impediment to doing that because one can multiply a unitary matrix by any
complex one. In our case matters are even simpler since the ``foreign'' matrices 
are diagonal with $U(1)$ elements on their diagonals, so they make group theoretical sense in terms of
$U(N)$. To be sure, all integration variables are $SU(N)$ matrices.  

Let the ordered diagonal matrices consisting of eigenvalues of $g=g_0 g_2....g_{N_t -1}$ at each spatial 
site ${\vec x}$ be:
\begin{equation}
	D_g({\vec x})={\rm diag}(e^{i\theta_1({\vec x})},e^{i\theta_2({\vec x})},....,
	e^{i\theta_N ({\vec x})}), \;\;\; \sum_{j=1}^N \theta_j({\vec x})=0 {\rm mod}(2\pi).
\end{equation}
We choose an $N_t^{\rm th}$ root of $D_g(x)$, $D_P(x)$,
\begin{equation}
	D_P(x)= {\rm diag} {\Big ( } e^{i\frac{\theta_1(x)}{N_t}},............,e^{i\frac{\theta_N(x)}{N_t}}{\Big )}
\end{equation}
By a further variable transformation (acting like an ordinary gauge transformations, but with phase elements not in $SU(N)$), each link in the time direction can be made to 
carry one copy of $D_P(x)$. Note that there are only $L^d$ $D_P({\vec x})$ variables; they do not depend on $t$. In the path integral one integrates over $L^d$ sets of $\theta({\vec x})$'s. They 
are nonlocal in $t$, constrained to be identical on all $N_t$ links in the $t$ direction.

\subsection{Finite temperature effective potential to one loop.}

The calculation in this subsection is a generalization of an old result~\cite{fintempnpb}. The presentation here will be 
more detailed.
The objective is to 
derive in leading order perturbation theory the effective potential governing the distribution of the eigenvalues of an untraced Polyakov loop in a manner similar to the 
derivation of the effective potential for all angles in the Eguchi Kawai model~\cite{qek}. 
By ``effective potential'' I mean the familiar term in the  generating function of one-particle-irreducible diagram. It is needed here for fields taking values in a curved compact space, the $SU(N)$ group manifold. The ultraviolet regularization is the discrete lattice and the infrared one comes by keeping only a finite number of sites. There is no ``large field problem''.

I work in space-dimension $d$ on a torus of $L^d \times N_t$ sites. The limit  $L\to\infty$ will
also be considered. The spatial directions are labeled by $\mu=1,2,...,d$. The (d+1)-direction is $\tau$. 
A site location on the $d+1$ dimensional lattice is labeled by $(x,t)$. $x$ has $d$ components and $t$ 
is the $d+1$ component of the site index. $\tau$ is the unit vector in the $d+1$ direction. 
The partition function, $Z$, up to normalizations which are allowed to change as the calculation progresses, is:  
\begin{equation}
	\begin{split}
		Z= &\int \prod_{x} \prod_{t=0}^{N_t -1} {\Big \{ } \prod_{\mu=1}^d [dU_\mu(x,t)] dU_0(x,t)  {\Big \}} e^{S(U_\mu , U_\tau)}\\
		S(U_\mu , U_\tau ) =&\frac{\beta}{2}\sum_x \sum_{t=0}^{N_t}{\Big \{}\sum_{1\le\mu,\nu\le d}
		tr [U_\mu(x,t) U_\nu (x+\mu ,t )U_\mu^\dagger (x+\nu ,t)U_\nu^\dagger (x,t)]+\\
		&c \sum_{\mu=1}^d tr[U_\mu(x,t) U_\tau (x+\mu, t )U_\mu^\dagger (x,t+\tau)U_\tau^\dagger(x,t)+h.c.]{\Big \}}\\
		=&-\frac{\beta}{2} {\big \{ }\sum_{1\le\mu < \nu\le d} Tr[T_\mu ,T_\nu][T_\mu,T_\nu]^\dagger + c \sum_{\mu=1}^d [ Tr[T_\mu , T_\tau]	[T_\mu, T_\tau]^\dagger +Tr[T_\mu , T_\tau^\dagger ]	[T_\mu, T_\tau^\dagger ]^\dagger ] {\big \}}
	\end{split}
\end{equation}

The gauge is completely fixed as before. There are $L^d$ angle variable sets, $\theta_j(x)\in[-\pi,\pi)$, $j=1,...,N$ ordered in increasing order in $j$. They do not depend on $t$. 
The effective potential for the angles $\theta$ is computed to leading order in $\frac{1}{\beta}$, 
at fixed ratio $\frac{\beta_t}{\beta} = c$. 

\begin{equation}
	\begin{split}
	Z=&{\rm Cst} \int\prod_x \prod_k [ d\theta_k (x) ] \prod_{1\le k < j\le N} [\sin^2 \frac{\theta_j (x) -\theta_k (x)}{2}]
		\prod_x [{\delta}_{{\rm mod}(2\pi)} (\sum_k \theta_k(x))]
		\\ & \int \prod_{\mu,x,t} [ dU_\mu (x,t) ] e^{S(U_\mu (x,t),U_\tau (x,t)=D_P(x))}
	\end{split}
\end{equation}

The $\sin^2$ term is a Jacobian resulting from changing integration variables. The original ones were $SU(N)$ matrices $W(x)$, one for spatial site $x$, the new ones are their eigenvalues. The Jacobian at each $x$ shows that they repel. The unitary diagonalizing matrices are absorbed as spatial gauge transformations. 

The finite temperature global $Z(N)$ symmetry acts by $\theta_j (x) \rightarrow \theta_j (x) +\frac{2k\pi}{N}, k=0,1,2,..,N-1$. If the $e^{i\theta_j}(x)$ 
phases consist of all distinct $N^{\rm th}$ roots of unity for each $x$, elements of 
$Z(N)$ cyclically rotate one angle into the next, indicating a preferred 
$Z(N)$ invariant state compatible with repulsion. At $\beta=0$, 
the $Z(N)$ is preserved by the vacuum. When $\beta$ becomes large, the
action must be taken into account. The following leading order calculation in $1/\beta$ shows 
that the Jacobian domination will be reversed when $\beta\to\infty$. Somewhere on the positive $\beta$
axis a $Z(N)$ breaking phase transition will occur.

The calculation proceeds on very similar lines as in~\cite{qek}. One expands around 
a classical configuration that maximizes $S$ and dominates at infinite $\beta$:
\begin{equation}
	U_\mu^{ij}(x,t)=\delta_{ij},\;\;\;\theta_i(x)=\theta_i^0\;\;\; i,j=1,...,N
\end{equation}
The corresponding $T$-parallel transporters are denoted by $T_\mu^0$ and $T_\tau^0(\theta)$. They mutually commute.
Dependence on $\theta_i^0$ enters only through $T_\tau^0(\theta)$. 

One needs to integrate over fluctuations $\delta\theta^j(x)$ and $\delta U_\mu (x,t)$ at quadratic order
in the exponents. They induce fluctuations $\delta T_\mu , \delta T_\tau$. Since the background is commutative and the action consists of a commutator times its adjoint and traced over, second order contributions arise solely from
first order variation of the commutators. The remaining $T$ factors are $T^0$'s.

There will be no quadratic cross terms containing $\delta T_\mu$ and $\delta T_\tau$: 
\begin{equation}
	Tr ( \delta T_\mu \delta T_\tau {T_\mu^0}^\dagger {T_\tau^0(\theta)}^\dagger ) + 
	Tr (\delta T_\mu \delta T_\tau^\dagger {T_\mu^0}^\dagger T_\tau^0(\theta) )=
	Tr[ \delta T_\mu (\delta T_\tau {T_\tau^0 (\theta)}^\dagger  +\delta T_\tau^\dagger T_\tau^0(\theta)) {T_\mu^0}^\dagger]=0,
\end{equation}
because, due to the $t$-independence of $U_\tau (x,t)$, $[T_\tau, T_\tau^0(\theta)]=0$ and $[\delta T_\tau , {T_\tau^0(\theta)}^\dagger]=0$. Also, $\delta (T_\tau T_\tau^\dagger )=0$. 

Contributions from the measure term are at lower order in $\frac{1}{\beta}$ because there was no $\beta$ dependence 
in the Jacobian to start with. 

The decoupling of $\delta T_t$ and $\delta T_\mu$ fluctuations can also be seen directly at the level of plaquettes, by cancellations of cross terms coming from $t$-consecutive time-like plaquettes. 

Integration over fluctuation of the $\delta T_\tau - \delta T_\tau$ type will make a contribution that is
independent of the $\theta_i^0$ because the $T_\tau^0(\theta)$ factors will cancel out. In summary,
for the purpose of computing the effective potential, we only need to integrate over fluctuations in $U_\mu$.
Two types of terms will be generated: one associated with terms coming from $[T_\mu, T_\nu]$ -- leaf plaquettes -- and the other from $[T_\mu, T_\tau]$  -- plaquettes in the $\tau$ direction. Fluctuations in $T_\mu$ that are diagonal in color indices can also be ignored because
they do not couple to the $\theta^0_i$; only off-diagonal fluctuation need to be taken into account.
To be sure, the integration over diagonal fluctuations is under control: they are explicitly compact. 

\subsubsection{Contribution from the $ [T_\mu, T_\tau] $ term.}

The quadratic contribution comes only from the first order variations $\delta F_{\mu,\tau}$ because 
the action is proportional to $Tr[ F_{\mu,\tau} F_{\mu,\tau}^\dagger ] $ and at zeroth order
all matrices commute. The variation in $T_\mu$ is
\begin{equation}
	\delta T_\mu = \delta (T_\mu^0 \mathbf{U}^\mu) = T_\mu^0 \delta \mathbf{U}^\mu=T_\mu^0 (\mathbf {1} + i\delta A_\mu)
	\end{equation} 
where $\delta A_\mu$ is hermitian. The second term in the commutator, $T_\tau (\theta )$, is a square matrix 
with $N N_t L^d$ columns whose eigenvalues are $L^d$-degenerate and given by 
$e^{\frac{2\pi i}{N_t} l +i\theta_j}, l=0,1,...,N_t - 1 ,j=1,...,N$. 
Conjugating by DFT $\Omega$ simultaneously for all $d+1$ directions, $\Omega^* T_\tau^0 \delta A_\mu \Omega \equiv\delta { \hat A}_\mu $, yields matrices circulant in all $d+1$ momentum indices. The $\delta A_\mu$ contain all the integration variables in matrix form. The integration measure 
is associated with the norm $Tr \delta A_\mu \delta{ A_\mu}^\dagger$. This measure is invariant under conjugation of $\delta A_\mu$ by a unitary matrix. We can integrate 
over $\delta { \hat A}_\mu $ instead of $\delta {A}_\mu $. The
leaf index $t$ goes into an index $l$ and the $d$-component site index $x$ goes into a $d$ component
momentum index $p$.

The basic structure is identical to the one in~\cite{qek}: Let $X$ be a complex matrix and $D$ a diagonal unitary one. Then $[X,D]_{a,b} =  X_{a,b}(D_b-D_a)$. With $D_a=e^{i\gamma_a}$, we find

\begin{equation}
Tr[X,D][X,D]^\dagger = 2\sum_{a<b} |X_{a,b}|^2 \sin^2 \frac{\gamma_a - \gamma_b}{2}
\end{equation}

This simple formula is used in the intermediate steps from equation (7) to (9) in~\cite{qek}. In our case $a,b$ are compound indices, $a=(t,x,j)$ and $b=(t',x',k)$.

\begin{align}
	\begin{split}
&Tr [\delta T_\mu , T_\tau^0(\theta)][\delta T_\mu , T_\tau^0 (\theta)]^\dagger =
2\sum_{l,p,l',p',k<j} |(\delta {\hat A_\mu})_{(l,p,j),(l',p',k)}|^2 \sin^2 \frac{2\pi (l-l') +\theta^0_j-\theta^0_k}{2N_t}\\&
=2 N_t L^d \sum_{l,p,k<j} |(\delta {\hat A_\mu})_{(0,0,j),(l,p,k)}|^2 \sin^2 \frac{2\pi l +\theta^0_j-\theta^0_k}{2N_t}
\end{split}
\end{align}

The angles $\theta^0$ only enter through the differences $\theta^0_i - \theta^0_j$. Hence, we can ignore contributions coming from diagonal color components $(\delta{\hat A}_\mu)_{(l,p,j),(l',p',j)}$ as they produce terms independent of $\theta^0$.

\subsubsection{Contribution from the $[T_\mu , T_\nu ]$ term.}

The angles $\theta^0_j$ do not enter directly, but the commutator contributes to similar terms in the action which do. In isolation, the $[T_\mu , T_\nu ]$ terms are $N_t$ decoupled $d$-dimensional Wilson action lattice theories. Hence, second order contribution must contain ``zero modes''  that correspond to a gauge transformation acting independently on leaves. The contribution evaluated before provides masses to all these modes. To be sure, the full system has no leaf-restricted gauge symmetry. 
The leafs are coupled through the angles: had we integrated over them fully, we would have returned to the projector on gauge invariant leaf states we discussed earlier. 

Let us pick one pair $\mu\ne\nu$. 

\begin{equation}
	\delta [T_\mu,T_\nu] =[\delta T_\mu , T_\nu^0]-[\delta T_\nu , T_\mu^0] 
\end{equation}

We need to evaluate the second order term $Tr[\delta[T_\mu,T_\nu](\delta[T_\mu,T_\nu])^\dagger$. The $\delta  {\hat T}_\mu$ -- $\delta {\hat T}^\dagger_\mu$ and $\delta {\hat T}_\nu$ -- $\delta {\hat T}^\dagger_\nu$ 
terms have familiar structures. The $\delta {\hat T}_\mu$ -- $\delta {\hat T}^\dagger_\nu$ 
and $\delta {\hat T}_\nu$ -- $\delta {\hat T}^\dagger_\mu$
come with a projector in the $d$-dimensional space indexed by $\mu,\nu$ and 
proportional to $\sin\frac{2\pi(p_\mu-{p'}_\mu)}{2L} \sin\frac{2\pi(p_\nu-{p'}_\nu)}{2L}$. 
This form is obtained after absorbing some momentum dependent phases in the $\delta{\hat A}_\mu$ variables, which drop out. The quadratic form connecting all $\delta{\hat A}_\mu$, $\delta{\hat A}_\nu^*$ variables
has zero modes. All of them get masses from the $\mu-\tau$ plaquettes. So, there are no divergences in 
the path integral, even when truncating to this order in $\beta^{-1}$.
If none of the momentum components are zero, one can choose a new basis of $d$ component vectors, different at
each momentum, which separates out the zero modes explicitly. There will be unresolved degeneracies when some or all
components of the momenta vanish. This is not a problem, one just needs to linearly complete the choices in some
way. Traditionally, in the continuum one ignores this, and goes to longitudinal and transverse components.
For notational simplicity I'll do this below, with the understanding that nothing is fudged even at these 
degeneracy points.

${\hat A}_\mu$, viewed as a $d$-vector, is decomposed into 
$d-1$ $d$-component vectors transverse to the vector $\sin\frac{\pi (p_\mu-p'_\mu)}{L}$,  $e^{(\alpha)}$, 
and a longitudinal $e^{\rm long}$. These vectors depend on $p-p'$. 
${\hat A}_\mu = \sum_{\alpha =1}^{d-1}{\hat A}^{(\alpha)} e^{(\alpha)}_\mu +{\hat A}^{\rm long} e^{\rm long}_\mu$. The $e^{(\alpha)}$ and $e^{\rm long}$ make up an orthonormal set under the inner product 
$\braket{e|e^{'}}=\sum_{\mu=1}^d e^{*}_\mu e^{'}_\mu$. The ${\hat A}^{(\alpha)}$ carry the same indices as ${\hat A}_\mu$ and so does ${\hat A}^{\rm long}$, which does show up explicitly in the term coming from spatial plaquettes. 

In summary, the total contribution to the action from spatial plaquettes, ignoring additive constants, is
\begin{equation}
Tr[T_\mu , T_\nu ][T_\mu , T_\nu ]^\dagger =	2 N_t L^d \sum_{1\le k,j \le N} \sum_{l,p} \sum_{\alpha=1}^{d-1}
	|( \delta {\hat A}^{(\alpha)}_{(0,0,j)(l,p,k)} |^2 [\sum_{\nu=1}^d \sin^2 \frac{2\pi p_\nu}{2L} ]
	\end{equation}

\subsubsection{Total contribution.}

The integration measure is 
$\prod_{\mu=1}^d\prod_{l=0}^{N_t -1}  \prod_{p} \prod_{1\le k<j
	 \le N} d{(\delta {\hat A}_\mu)}_{(0,0,j),(l,p,k)} d[{(\delta {\hat A}_\mu)^{*}}_{(0,0,j),(l,p,k)}]$. 
One can replace $\mu$ and the associated $\mu$ indexed variables by the set $\alpha$ and ``${\rm long}$'',  
because the transformation relating them is orthonormal.  
(The diagonal $j=k$ terms would be real, but they can be ignored as explained.)
This expression
contains the list of all independent (complex) variables which are integrated over. 
The integrand is factorized into elementary Gaussian terms. Each independent 
integration variable can be rescaled without altering the $\theta^0$ dependent contribution 
to the effective potential. This changes an overall normalization which is irrelevant for the probability distribution of the angles $\theta^0$. With $c^{-1} = \rho$ the integrand's exponent is:
\begin{align}
	\begin{split}
	&-L^d N_t  \sum_{1\le k<j\le N}  {\sum}_{l,p}  [ \sum_{\alpha=1}^{d-1} |(\delta {\hat A}^{(\alpha)})_{(0,0,j),(l,p,k)}|^2 
		[ \sin^2 \frac{2\pi l+\theta^0_j-\theta^0_k}{2N_t}  +\rho\sum_{\nu=1}^d \sin^2\frac{2\pi p_\nu}{2L} ]+\\&
		|(\delta {\hat A}_\mu^{\rm long})_{(0,0,j),(l,p,k)}|^2 \sin^2 \frac{2\pi l+\theta^0_j-\theta^0_k}{2N_t}]
		\end{split}
\end{align}

Above, ${\sum}_{p}$ is over all $p$, in ranges $0,..,L-1$ for each component. 
The constant $\rho>0$ is of order $1$. The $\theta^0_j$ integration measure factor needs to be added to $v(\theta^0)$, 
the result obtained by integration over fluctuations.

This expression can be understood as follows: We are working in an axial gauge
corrected for compactness in the direction that was gauge fixed, to 
leading order in perturbation theory around a configurations that
breaks $d+1$ gauge invariance. This correction endows
the transverse gauge field components  with distinct 
masses, explicitly depending on group indices. In this deconstructed $d$-dimensional framework 
we have $\frac{N(N-1)}{2} N_t$ massive gauge bosons represented by 
as many complex fields. There also are $(N-1)N_t$ real bosons whose masses 
are not dependent on the color index. $N-1$ out of the latter have zero mass.

 The term $\sin^2 \frac{2\pi l+\theta^0_j-\theta^0_k}{2N_t}$ above, appearing in sums over the discrete momenta in the $\tau$ direction, indicates how fluctuating values of the ordered $\theta$ angles fill in the gaps of length $\frac{2\pi}{N_t}$ created by the integral values of $l$, reflecting finite temperature and
 periodicity in the $\tau$ direction. If the spacings are uniformly filled, the dependence on temperature 
 disappears. At infinite $N$ the distribution of eigenvalues fills the gaps continuously for large
 enough (but finite) $N_t$. If the global $Z(N)$ is preserved the distribution is flat and the fill is ``perfect'' for any $N_t$. 
     
  Before carrying out the integral over fluctuations we wish to eliminate the products over $l$. This
  leaves one with a purely $d$-dimensional point of view, as desired in a deconstruction. 
     
   The sum over $l$ can be performed starting from the identity (known to Mathematica):
   \begin{align}
    \begin{split}
     2^{n-1}&	\prod_{k=0}^{n-1} [\cos x -\cos ( y+\frac{2k \pi}{n}) ] = \cos nx - \cos ny\\
     2^{2n}&\prod_{k=0}^{n-1}
     [\sin^2(\frac{y}{2}+\frac{k\pi}{n})-\sin^2\frac{x}{2}] =\sin^2\frac{ny}{2} - \sin^2\frac{nx}{2}
    \end{split}
    \end{align}
  This identity is between polynomials in $e^{ix}$ and $e^{iy}$. It can be proved
  by matching zeros and normalizations. Let us first focus on the $\delta {\hat A}_\mu^{\rm long}$ 
  term. Using the second form of the identity the integration over the longitudinal component produces
  a factor
  \begin{equation}
   \prod_p \prod_l \prod_{k<j}\prod_{l=0}^{N_t} \sin^2 (\frac{\theta_j-\theta^0_k}{2} +\frac{\pi l}{N_t})=
    {\rm Cst} \prod_{k<j} (\sin^2 \frac{\theta_k-\theta_j}{2})^{L^d}
   \end{equation}
   in the denominator. This exactly cancels the Jacobian  factor evaluated at the uniform configuration
   $\theta_j(x)=\theta^0_j$. The upshot is that only the transverse components on the spatial leaves 
   contribute to the effective potential. 
   Here, this is a consequence
   of combining restricted gauge invariance with the projection 
   on gauge singlets. 
     
   We now turn to this remaining contribution of the transverse components of the 
   vector potential on the spatial leaves. Replacing $x$ by $ix$ in the above identity we get
   \begin{align}
   \begin{split}
   	2^{n-1}	\prod_{k=0}^{n-1} [\cosh x -\cos ( y+\frac{2k \pi}{n}) ] =& \cosh nx - \cos ny \\
   	2^{2n}	\prod_{k=0}^{n-1} [\sin^2 ( \frac{y}{2}+\frac{k \pi}{n}) + \sinh^2\frac{x}{2}]=&
   	\sin^2\frac{ny}{2}+\sinh^2\frac{nx}{2} 
   \end{split}
   \end{align}

   To use this variant of the identity we define the quantity $R(p)$ by:
   \begin{equation}
   	R (p) = \cosh^{-1} [1+\rho\sum_{\nu} \sin^2\frac{\pi p_\nu}{L} ],\;\; R(p) \ge 1.
   	\end{equation}

   Then,
   \begin{equation}	
   		v(\theta^0 ) =
   		-(d-1)\sum_{1\le i < j \le N} {\sum}_{p} \log (\sin^2 \frac{\theta^0_i-\theta^0_j}{2}
   		+\frac{\cosh(N_t R(p))-1}{2}).
   \end{equation} 

Using the definition of Chebyshev polynomials for arguments outside the interval $(-1,1)$,
\begin{equation}
	T_n(\cosh x)= \cosh(n x),
\end{equation}
we obtain: 
\begin{equation}	
		v(\theta^0 ) =
		-(d-1)\sum_{1\le i < j \le N} {\sum}_{p} \log [\sin^2 \frac{\theta^0_i-\theta^0_j}{2}
		+\frac{T_{N_t} (1+\rho\sum_{\nu=1}^{d}\sin^2\frac{\pi p_\nu}{L})-1}{2}].	
	\end{equation}
Discarding another constant, the formula becomes
\begin{equation}	
	v(\theta^0 ) =
	-(d-1)\sum_{1\le i < j \le N} {\sum}_{p} \log [T_{N_t} (1+\rho\sum_{\nu=1}^{d}\sin^2\frac{\pi p_\nu}{L})-\cos(\theta^0_j - \theta^0_i )].	
\end{equation}

The effective $V_{\rm eff} (\theta^0 )$ potential is defined per unit spatial volume with a minus 
sign so that its minimum gives the vacuum energy density. We discard an $N_t$ 
dependent constant which carries no $\theta^0$ dependence. 
The discarded term is a temperature dependent vacuum energy correction.
\begin{equation}
	V_{\rm eff} (\theta^0 ) =\sum_{1\le i < j\le N} {\mathfrak{e}} (\theta^0_j -\theta^0_i ),\;\;\;
	{\mathfrak{e}}(\gamma )=\frac{d-1}{L^d} \sum_p \log [1-\frac{\cos\gamma}{T_{N_t} (1+\rho\sum_{\nu=1}^{d}\sin^2\frac{\pi p_\nu}{L})}]
\end{equation}

It is now obvious that $\gamma=0$ is a minimum of $\mathfrak{e}(\gamma)$ showing that the angles $\theta^0_i$ now attract and could therefore cause spontaneous $Z(N)$ symmetry breakdown. 

The formula holds at finite $N$, $N_t$, $L$ and lattice spacing (set to 1, but easily formally replaced by ``$a$'').
The $L\to\infty, N_t\to\infty , N\to\infty , a\to 0$  limits can be taken in different orders or various correlated ways. Exploring this systematically  would take us too
far afield. As an example, I shall take the infinite $L$ limit followed by the formal ``$a\to 0$'' limit,
taken only in the spatial directions. This is the ``deconstruction'' limit. It amounts to $\rho\to\infty$, 
because the spatial lattice spacing becomes infinitely smaller than the separation among leaves. 
The infinite $L$ limit (assuming it converges) is:
\begin{equation}
{\mathfrak{e}}^{L=\infty} (\gamma )=(d-1) \int_{|k_\mu|<\pi}\frac{d^d k}{(2\pi)^d} \log [1-\frac{\cos\gamma}{T_{N_t} (1+\rho\sum_{\nu=1}^{d}\sin^2\frac{k_\nu^2}{2})}]
\end{equation}
A finite continuum limit requires $N_t > \frac{d}{2}$. Thinking about 
momenta in units of mass, we need to give $\rho$ inverse mass squared units. 
Then, using a mass dimensional parameter to measure all momenta we take the 
$\rho\to\infty$ limit, expressing everything with the help of one mass scale, $f$:
\begin{equation}
[{\mathfrak{e}}^{L=\infty}]_{\rho=\infty} (\gamma )=\frac{(d-1) f^d}{\Gamma(\frac{d}{2}) (2\pi)^{\frac{d}{2}}} \int_{x=0}^\infty x^{d-1}dx  \log [1-\frac{\cos\gamma}{T_{N_t} (1+x^2)}]
	\end{equation}
    
The Chebyshev polynomial in the denominator increases rapidly with $x^2$ since its argument is outside $[-1,1]$, so expanding
the logarithm to leading order captures the relevant qualitative properties for reasonable values of $N_t$ and $\cos \gamma$.

While now $\cos\gamma$ goes to 1 at the minimum of the effective potential, and a $Z(N)$ symmetric probability distribution
for the $\theta^0_j$ variables is disfavored, one needs to keep in mind that the formula 
only holds as a leading order computation. In a simulation, one should not expect to see 
degeneracies of the $\theta_0^j$: the repulsion term in the measure is exact, and therefore there always will be peaks
of a width $\sim\frac{1}{N}$ centered at $N^{\rm th}$ roots of unity. The height of peaks would be decreasing away
from the central peak at $\gamma=0$ in both directions for $\gamma\in[-\pi, \pi]$.  For large enough $\beta$ the $Z(N)$ 
would be spontaneously broken. One could extend the calculation to include a spatial angle-derivative term to allow
consideration of angles varying slowly on the leaves. 

To get at the standard continuum finite temperature effective potential the simplest would be to
keep $\rho=1$ and take $N_t$ and $L$ to infinity in a correlated way, producing a finite 
temperature and an infinite spatial volume in the end. The effective potential here is 
deconstructed and different. 

The $N_t=1$ case violates the bound $N_t >\frac{d}{2}$ at the physical point $d=3$, which blocks taking continuum limit in the spatial directions. One still can take $L\to\infty$. 
The continuum limit for the effective potential can be taken after appropriate subtractions. 
In this superrenormalizable case one has good control over the ultraviolet. One has a gauge fixed three dimensional $SU(N)$ gauge theory interacting with a compact, $SU(N)$-valued  Higgs field in the adjoint.

\section{The {\"U}Y deformation.}
      
     In the last equation of~\cite{qek} (eq.(18)) it was pointed out that the quantity
     $P(\theta)=\frac{1}{N^2} \sum_{i,j=1}^N \sin^2\frac{\theta_i-\theta_j}{2}$ is bounded from above by 
     $\frac{1}{2}$ for any set of angles and that for a uniform distribution the bound was saturated. In the numerical simulations of that paper the departure from saturation was used to estimate the location of $Z(N)$ symmetry breaking transitions. The bound is easy to prove:
     \begin{equation}
     	\frac{1}{N^2}|\sum_{i=1}^N e^{i\theta_i}|^2 =1-2P(\theta)
     	\end{equation}
     It is not true that saturation of the bound implies uniformity: If $N=mn$ with $m,n$ integers
     larger than 1, the index $i$ can be replaced by a double index $(a,b)$ with $1\le a \le m$ and $1\le b\le n$. In a manipulation familiar from FFT (fast Fourier transform), the sum
     over $i$ can be reorganized
     \begin{equation}
      \sum_{i=1}^N e^{i\theta_i}= \sum_{b=1}^n \sum_{a=1}^m e^{i\theta_{a,b}}
      \end{equation}
     To saturate the bound it is enough that the sum over $a$ vanish 
     for all $b$. Such an angle set could preserve only a $Z(m)$ subgroup of $Z(mn)$: 
     the sum over $n$ at fixed $m$ does not need to vanish. 
     It is for this reason that all recent large $N$ simulations that relied
     on a qualitative large $N$ suppression of finite volume effects I have been involved in have been restricted to prime $N$.
 
     In its simplest form, the {\"U}Y deformation is a new term in the action given 
     by a sum over spatial sites  $\sum_{x} P(\theta(x))$ (up to an irrelevant constant).  
     It changes the dependence on time link variables
     of the action itself, not just of some observables. When only the latter are impacted, 
     the change is limited. The observables can be viewed as probes of the theory's 
     vacuum. They act as a limited number 
     of excitations over a homogeneous vacuum, similar to particle states (in this case allowing also
     configurations extended in space, like strings).  
     
     The new term is taken with a sign promoting angle repulsion. It counteracts the cancellation 
     of the Jacobian by fluctuations in the spatial link variables and, beyond a certain strength, 
     will restore $Z(N)$ symmetry. In this phase the system reconfines~\cite{delia}. 
     
\subsection{The \"{U}Y term biases gauge averaging.}
          
 I now turn back to the derivation of a transfer matrix and the associated projector
 on gauge invariant states starting from the original action. It is much simpler 
  to first fix the gauge such that all time like link variables are fixed to unity
  except one, whose link variable is the open Polyakov loop. 
      
  Obviously, the deformation affects gauge averaging. With the term present
  the derivation of the transfer matrix will not produce a projector on singlet
  states because the measure in the averaging procedure is changed from Haar by
  a weight dependent on the character of the adjoint representation of $SU(N)$. 
  This would still prohibit $Z(N)$ charge carrying representations but would 
  allow all $SU(N)/Z(N)$ representations (in other words, only the $Z(N)$ subgroup of $SU(N)$ still is averaged over). The projectors we obtained in the 
  undeformed theory get replaced by an operator $\mathfrak{Y}$, which is not a projector. 
  For example, a spatial open (untraced) Wilson loop will have a deformation dependent expectation value, while it would vanish in the absence of deformation. 
  The deformed gauge averaging operator $\mathfrak{Y}$ is inserted once for each site index $x$. Denoting the extra term in the action by a $Z(N)$ invariant function of a group element, $\mathfrak{U}(g)$, we find: 
  \begin{equation}
    \mathfrak{Y}=\int [dg] e^{\mathfrak{U}(g)}\sum_\mathfrak{R} \Psi_\mathfrak{R} (V_1) \overline{\Psi_\mathfrak{R} (^{g}V_2)}
  \end{equation}
   The action is:
   \begin{equation}
   \mathfrak{Y} \Psi (V) = \Psi_{Y} (V) = \int [dg] 
   e^{\mathfrak{U}(g)} \Psi(^{g}V)
   \end{equation}
   Convolving once we get
   \begin{equation}
    [\mathfrak{Y*Y}] \Psi (U) = \Psi_{Y*Y} (\mathfrak{U}) = \int [dg'] e^{\mathfrak{U}(g')} \int [dg] e^{\mathfrak{U}(g)} 
    \Psi(^{g'}{^{{g}}U})
    \end{equation}
    $\mathfrak{Y*Y}$ clearly is not the same as $\mathfrak{Y}$: it is not a projector. Each application applies the same weighted average and there is a cumulative effect to multiple applications.  When $\mathfrak{Y}$ acts on an overall gauge singlet state, it gets reproduced times some constant. In other words, up to a (calculable) multiplicative constant. $\mathfrak{Y}$ is a projector if acting on an eigenstate of $\mathfrak{P}$.

       The {\"U}Y deformation alters the entire Hilbert space of the undeformed theory. 
      So long as the system is finite in all directions 
      the distinction is only qualitative and one can imagine how the vacuum gets deformed
      adiabatically as the deformation is slowly turned on.
      This assumes that the infinite sums over representations converge. At infinite volume though, 
      the deformed Hilbert space becomes orthogonal to the undeformed one. Something similar
      will happen also when one takes $N\to\infty$, while the system size stays finite.

      Intuitively, the fraction of states propagating with the $\mathfrak{Y}$ insertion and that are 
      gauge singlets is small because there are infinitely more states carrying all possible $SU(N)/Z(N)$ 
      irreducible representations of the gauge group. So far, I have not been able to turn this into
      a conclusion that in a standard Monte Carlo simulation the reconfined results 
      obtained with the {\" U}Y deformation one obtains are overwhelmingly
      unrelated to those obtained in the absence of the deformation. 
      
      If we follow $\mathfrak{Y}$ immediately by the projector $\mathfrak{P}$, only the singlets remain. In this case the deformation decouples and we can integrate it out. This result may be of some use in the
      future, but does not close the issue. 
      
      The one time link partition function for the simplest form of the deformation is given by:
      \begin{equation}
      {\cal J}(h)=\prod_{i=1}^N [\int_{-\pi \le \theta_i <\pi} \frac{d\theta_i}{2\pi} ]
      {\prod_{i<j}} [ e^{2h \cos(\theta_i - \theta_j)}\sin^2\frac{\theta_i-\theta_j}{2}]
      \end{equation}
      Expand each factor in a Fourier series:
      \begin{equation}
      \begin{split}
      &e^{2h\cos(\phi)}\sin^2(\frac{\phi}{2})=\frac{1}{2} p_0+\sum_{n=1}^\infty p_n \cos(n\phi)\;\;\;with\;\;\;
      p_n = \frac{1}{\pi} \int_{-\pi}^\pi d\phi e^{2h\cos(\phi)}
      [\frac{1-\cos(\phi)}{2}] \cos (n\phi)=\\
      &\frac{1}{\pi} \int_0^\pi d\phi e^{2h\cos(\phi)} [\cos(n\phi) -
      \frac{\cos((n+1)\phi)+\cos((n-1)\phi)}{2} ] = \\&I_n (2h) -\frac{I_{n+1}(2h)+I_{n-1}(2h)}{2}\;\;\;\;\;\;with\;\;\;\;\;\; I_k(z)\equiv\frac{1}{\pi}\int_0^\pi e^{z\cos\theta}\cos(k\theta)=I_{-k}(z)
      \end{split}
      \end{equation}
      Inserting the Fourier series in each factor of the integral defining  
       ${\cal J}(2h)$, we get
       \begin{equation}
       	\begin{split}
       {\cal J}(h)=&\prod_{i=1}^N [\int_{-\pi \le \theta_i <\pi} \frac{d\theta_i}{2\pi} ]\\
       {\prod_{i<j}}&{\big\{} \frac{1}{2} [I_0(2h)-I_1(2h)]+ [\sum_{k_{ij} =1}^\infty {\big [} I_{k_{ij}} (2h) -\frac{I_{k_{ij}+1}(2h)+I_{k_{ij}-1}(2h)}{2}{\big ]} e^{{ik_{ij} (\theta_i-\theta_j)}}\big{\}}
       \end{split}
       \end{equation}
      Now consider all terms containing at least one 
      $\theta_1$ in the exponent. To get a nonzero answer 
      when carrying out the $\theta_1$ integration one would need them
      to cancel each other, but the $k_{ij}\ge 1$ for all $i<j$. Continuing to the $\theta_2$ integral eliminates all $k_{2j}$ terms, and so on. The expression collapses to:
      \begin{equation}
      \begin{split}
      {\cal J}(2h)=&{\big [} \frac{I_0 (2h) -I_1(2h)}{2}{\big ]}^{\frac{N(N-1)}{2}}{\rm{~with}}\\
      {\cal J}(0)=&\frac{1}{2}\;\;\;{\rm{because}}\;\;\; I_0(0)=1\;\;{\rm{and}}\;\; I_1(0)=0\\
      \frac{{\cal J}(2h)}{{\cal J}(0)}=&[ I_0 (2h)-I_1(2h)]^{\frac{N(N-1)}{2}}
      \end{split}
      \end{equation}
      For all $x>0$, $1>I_0(x)-I_1(x) > 0$, while  $I_0(x)-I_1(x) > 1$ for all $x<0$. So,
      ${\cal{J}}(2h) >0$ for all $h$.
      At $N=\infty$, $\frac{{\cal J}(2h)}{{\cal J}(0)}$ is $+\infty$ when $x<0$ and $0$ 
      at $x>0$.       
      ${\cal J} (2h)$ is more properly viewed as an integral over the 
      group $SU(N)/Z(N)$ as nothing changes under the shift $\theta_i \rightarrow \theta_i+ r$ for all $i$ where $e^{2i\pi{r}}$ is an $N$-th root of unity. Since we normalized group integration to unity for $SU(N)$ and only look at ratios, this makes not difference. 
      
      In summary, the fraction of gauge invariant states relatively to the total number of states
      allowed for $h > 0$ (the relevant range for reconfinement) is suppressed by
      a power of spatial volume and $N(N-1)/2$. When either or both are infinite, the theory seems
      very different from ordinary, all scale confining pure $SU(N)$ gauge theory. It is the
      latter that is of interest in the quest of total large $N$ reduction.
      
      In~\cite{UY} it is argued that in the absence of $Z(N)$ breaking the Schwinger
      Dyson equations at leading order in $N$ remain independent of $h$. I see nothing wrong with the formal argument, nor with the claim that a $h$ of right size will restore global $Z(N)$. At strong couplings, $\beta, \beta_t,h$ small, this argument ought to be convincing. The convergence radius of the strong coupling series for $h$ (the 
      usual counting rules say $h$ is not scaled by any power of $N$) may be just 
      smaller than the value needed for reconfinement. For larger $h$ the dramatic change 
      in the Hilbert space structure takes the system to a phase different from standard QCD in a major way. Strictly speaking, the lattice Schwinger-Dyson loop equations loose their validity outside the strong coupling regime. Here it seems that it is a phase transition
      that blocks their validity. I suspect that at infinite $N$ the reconfined regime of 
      finite temperature pure gauge theory describes something very remote from ordinary
      finite temperature QCD and therefore the generalization meant to solve the 
      problem of full one site large $N$ reduction will not work.
      
      It also should be kept in mind that large $N$ reduction does work in continuum without any deformations, for box sizes larger than about $\frac{1}{T_c}$ where $T_c$ is the finite temperature transition in pure gauge theory in terms of the string tension at infinite volume~\cite{contek}. There have been numerical tests of the \"{U}Y deformation for $N=3,4$ and
      temperatures above $T_c$ by some modest amount and it was found that some physical quantities seem consistent with an adiabatic deformation of the same in the undeformed 
      theory, taken at temperatures in the regular confined phase. For one site reduction
      to work automatically in the deformed theory one would need to see results at larger
      values of $N$ and an analysis of their approach to the $N=\infty$ limit. 
      
      When testing whether \"{U}Y is an adiabatic deformation at large $N$ 
      it should also be kept in mind that there is a lattice bulk transition at large $N$ in Wilson
      lattice pure $SU(N)$ gauge theory that invalidates strong coupling results before one
      gets into the perturbative regime. That bulk transition has to be reproduced at the 
      same coupling under any prescription that seeks to achieve single site reduction on the lattice at infinite $N$. Any prescription for single site reduction if it holds, has to do
      so on the lattice.

 \section{Reconfinement by quenching.}
 
 Getting back to finite temperature, I ask whether there is a quenching alternative to
 deformation reconfinement. 
 
 Using the fully gauge fixed form above it is natural to ask what happens if we quench the
 $\theta_i (x)$ variables. The same arguments that applied in the Eguchi-Kawai case would apply
 also here at large $N$. The angles are gauge invariant variables and there are only $N-1$ of
 them per site. This would support temperature independence in the E-K sense. 
 
 It might be of interest to consider this quenched deconstructed model and do simulations
 directly in the axial gauge. The split into two stages one of doing the gauge field integral at fixed angles, and the next of averaging the averages of the first step over the angles distributed with the Jacobian measure might make it easier to separate out physically relevant
 effects, and at the same time, hopefully have a more clear understanding at the Feynman planar diagram level of perturbation theory~\cite{grosskitazawa}. In ordinary, 
 not deconstructed finite temperature QCD, perturbation theory gets stuck at some order because
 of infrared divergences. In the deconstructed case, the theory is strictly three dimensional 
 and the infrared divergences definitely don't go away but ultraviolet divergences are simpler. 
 
 At the technical level, I prefer the deconstructed view: the angle repulsion term 
 comes in explicitly, because one never works with vector potentials in the Lie Algebra
 for the time direction. Taking the continuum limit in all four directions makes it awkward to 
 keep the repulsion term, the remnant of gauge averaging which is crucial 
 in deriving a Hilbert space view of what the standard periodic gauge invariant lattice 
 path integral tells us. Gauge averaging is a procedure well defined for compact groups. 
 In the continuum it may get mixed up with ordinary gauge fixing and the manipulations seem to
 me less reliable. 
 
 Local deformations to an action are typically expressed in terms of the same elementary fields.
 As a start, their renormalization can be borrowed from the renormalization of the observable corresponding
 to the undeformed theory which is well understood. In four dimensions the Polyakov loop is a composite  
 nonlocal operator, which requires non-perturbative renormalization. The single gauge invariant method I know of is by continuum smearing~\cite{smear} which comes with a dimensional parameter determined in the
 undeformed case when the operator is just an observable. When the latter is promoted to a term in the
 action it isn't even clear to me what the dimensions of the coupling constant $h$ should be. 
 
 So long as one relies on some undiscriminating version of effective field theory I think the 
 deconstructed framework is better suited, because then the operator is not really nonlocal.
 
 \section{Summary}
 
 My presentation has been very detailed. Only basic information about lattice gauge theory would be 
 required to understand the paper. I have not produced hard answers to the questions that motivated me. 
 I can only hope that there are some new insights here that would be useful in the future.
 
 The main points were:
 
 (1) The periodic Euclidean time lattice (path) integral of Wilson lattice gauge theory
 produces a transfer matrix acting in a Hilbert space consisting of square integrable 
 functions of spatial link variable configurations which are singlets under a local
 spatial gauge group. 
 
 (2) There is an exactly equivalent fixed gauge integral construction which has, in addition 
 to the spatial link variables, $N$ dynamical angles per spatial site and a repulsion term of a simple form dictated by the structure of $SU(N)$.
 
 (3) The \"{U}Y deformation adds an additional, angle dependent, term which alters the
 structure of the Hilbert space by allowing all $Z(N)$ invariant representations of the
 gauge group.
 
 (4) In the limit of $N\to\infty$ the entire \"{U}Y Hilbert space is orthogonal to the Hilbert
 space in the absence of the deformation.
 
 (5) There is a quenched alternative to the deformation, which also would restore $Z(N)$
 symmetry. 
 
 (6) The \"{U}Y deformation and the quenched version admit generalizations to all directions
 and both were proposed as ways to define a single site matrix model equivalent to the full
 Wilson pure gauge model. I believe that the quenched version has a better chance to work than
 the deformation version, but neither has been proven. In the continuum, on a torus
 of sides larger than the inverse deconfinement temperature large $N$ reduction holds 
 without either alteration. I don't think this is known for the \"{U}Y case, and am 
 skeptical that it would hold for the same continuum $SU(N)$ theory that the ordinary
 Wilson action produces.
 
 (7) It was speculated that starting from a deconstructed finite temperature $SU(N)$ gauge
 theory might be beneficial in that four dimensional ultraviolet behavior is not fully 
 turned on. 
 
 \section{Suspicions.}
 
I am skeptical about the {\" U}Y prescription for single lattice site reduction. 
The deformation is not adiabatic even for smallest values of
the its coupling. I cannot rule out being proven wrong by a thorough numerical simulation.

The root of the problem is that the deformation is nonlocal in one direction in the physical finite temperature system. The simultaneous addition of {\" U}Y deformations for all directions descends, from a system of $L^3 N_t$ sites arranged in a four dimensional hypercube, so I doubt it would work for extreme large $N$ reduction. 

All arguments about large $N$ stem from simple degrees of freedom counting.
It is this way of thinking that, seems to me, relies fundamentally on having a local action. Even when perturbation theories of different types 
organized themselves so that they admit large $N$ limits, conclusions
drawn about equivalences implicitly presuppose a unique vacuum. The way these conclusions can fail comes by a nonanalytic change in the 
``vacuum''. Even the restricted {\" U}Y deformation to finite temperature 
seems subject to the same worry. Again, I cannot rule out being proven wrong by a thorough numerical simulation.

Thorough numerical simulations are time and effort consuming and do not work in one blow, rather, require projects extending years. I am not sure 
that this problem will be deemed of sufficient interest to get embarked on by some individual or group.

There exists also another motivation for the \"{U}Y deformation, by introducing adjoint massless matter. This
theory is supersymmetric. Then, there is no $Z(N)$ breaking and Eguchi-Kawai 
reduction formally holds~\cite{susy} -- in the continuum limit. 
At one loop, an easy computation of massless adjoint fermion 
feedback gives a contribution exactly 
canceling the bosonic one, preserving the partition function = 1 rule in supersymmetry. This computation explicitly shows how 
the $Z(N)$ breaking is avoided in Feynman perturbation theory. 
One can break supersymmetry and 
give the fermions a large mass, so that the effective low energy theory does not see the supersymmetry but hope that large $N$ reduction is maintain at all volumes. 

On the lattice one would need the overlap~\cite{overlap} to permit
an emergent supersymmetry in the continuum because masslessness is protected. 
In this arrangement decoupling the adjoints by a large mass will take them
out of the game completely and the hope is dashed.

	\begin{acknowledgments}
		I thank Massimo D'Elia, Tony Gonzalez-Arroyo and Rajamani Narayanan  for comments.
	\end{acknowledgments}

	\section{References Cited}
	
	\clearpage
	

\begin{thebibliography}{99}
		\bibitem{brs2} H. Neuberger, Phys. Lett. B, 183, 337 (1987).
		\bibitem{MoMubook} ``Quantum Fields on a Lattice'' by Istv{\'a}n Montvay and Gernot M{\"u}nster. 
		\bibitem{UY} \"{U}nsal and L. G. Yaffe, JHEP 1008, 030 (2010).
		\bibitem{ek} Tohru Eguchi and Hikaru Kawai, Phys. Rev. Lett. 48 (1982) 1063.
		\bibitem{qek} Gyan Bhanot, Urs M. Heller, H. Neuberger, Phys Lett. B 113 (1982) 47.
		\bibitem{sw} B. Sheikholeslami and R, Wohlert, Nucl. Phys. B259 (1985) 572-596.	
		\bibitem{bm}  Y.V. Fyodorov and A.D. Mirlin, Phys. Rev. Lett. 67 (1991), 2405?2409.
		\bibitem{twist} A. Gonzalez-Arroyo and M. Okawa, Phys. Lett. B 120 (1983) 174; Phys. Rev. D 27 (1983) 2397.
		\bibitem{hagedorn} B. Hagedorn, Nuovo Cimento Suppl. 3, 147 (1965).
		\bibitem{ogil}  Michael Ogilvie, Phys.Rev.Lett. 52 (1984) 1369.
		\bibitem{fintempnpb} H. Neuberger, Nucl. Phys. B220[FS8] (1983) 234.
		\bibitem{deconstruction} Nima Arkani-Hamed, Andrew G.Cohen, Howard Georgi, Phys. Lett. B 513 (2001) 232.
		\bibitem{contek} J. Kiskis, R. Narayanan, H. Neuberger, Phys. Lett. B 574 (2003) 65.	
		\bibitem{delia} Claudio Bonati, Marco Cardinali, Massimo D'Elia, Matteo Giordano, and Fabrizio Mazziotti, Phys. Rev. D 103, 034506 (2021).
		\bibitem{grosskitazawa} David Gross, Yoshihisa Kitazawa, Nucl.Phys.B 206 (1982) 440-472
		\bibitem{smear} Rajamani Narayanan and Herbert Neuberger JHEP03 (2006) 064; Herbert Neuberger, 
		Phys. Rev. D 87, 114509 (2013).
		\bibitem{susy} R. L. Mkrtchyan and S. B. Khokhlach\"{e}v, Pis'ma Zh. Eksp. Teor. Fiz. 37, No 3, 160 (1983).
		\bibitem{overlap} Herbert Neuberger, Phys. Rev. D57 (1998) 5417-543

	\end{thebibliography}
\end{document}